\documentclass[letterpaper,aps,prb,preprint]{revtex4-1}
\usepackage{graphicx}
\usepackage{epsfig}
\usepackage{dcolumn}
\usepackage{bm}
\usepackage{times}
\usepackage{mathptmx}
\usepackage{color}
\usepackage{amsmath}
\usepackage{xfrac}
\usepackage{amsfonts}
\usepackage{float}

\begin{document}
	
\title{The rate of quasiparticle recombination probes the onset of coherence in cuprate superconductors}

\author{J.~P.~Hinton}
\affiliation{Materials Science Division, Lawrence Berkeley National Laboratory, Berkeley, California 94720, USA}
\affiliation{Department of Physics, University of California, Berkeley, California 94720, USA}
\affiliation{Department of Physics, University of California, San Diego, California 92093, USA}

\author{E.~Thewalt}
\affiliation{Materials Science Division, Lawrence Berkeley National Laboratory, Berkeley, California 94720, USA}
\affiliation{Department of Physics, University of California, Berkeley, California 94720, USA}

\author{Z.~Alpichshev}
\affiliation{Department of Physics, Massachusetts Institute of Technology, Cambridge, MA 02139}

\author{F.~Mahmood}
\affiliation{Department of Physics, Massachusetts Institute of Technology, Cambridge, MA 02139}

\author{J.~D.~Koralek}
\affiliation{Materials Science Division, Lawrence Berkeley National Laboratory, Berkeley, California 94720, USA}

\author{M.~K.~Chan}
\affiliation{School of Physics and Astronomy, University of Minnesota, Minneapolis, Minnesota 55455, USA}

\author{M.J.~Veit}
\affiliation{School of Physics and Astronomy, University of Minnesota, Minneapolis, Minnesota 55455, USA}

\author{C.J.~Dorow}
\affiliation{School of Physics and Astronomy, University of Minnesota, Minneapolis, Minnesota 55455, USA}

\author{N.~Bari{\v{s}}i{\'c}}
\affiliation{Institute of Solid State Physics, TU Wien, 1040 Vienna, Austria}
\affiliation{School of Physics and Astronomy, University of Minnesota, Minneapolis, Minnesota 55455, USA}

\author{A.~F.~Kemper}
\affiliation{Materials Science Division, Lawrence Berkeley National Laboratory, Berkeley, California 94720, USA}

\author{D.~A.~Bonn}
\affiliation{Department of Physics and Astronomy, University of British Columbia, Vancouver, Canada.}
\affiliation{Canadian Inst. Adv. Res., Toronto, ON M5G 178, Canada}

\author{W.~N.~Hardy}
\affiliation{Department of Physics and Astronomy, University of British Columbia, Vancouver, Canada.}
\affiliation{Canadian Inst. Adv. Res., Toronto, ON M5G 178, Canada}

\author{Ruixing~Liang}
\affiliation{Department of Physics and Astronomy, University of British Columbia, Vancouver, Canada.}
\affiliation{Canadian Inst. Adv. Res., Toronto, ON M5G 178, Canada}

\author{N.~Gedik}
\affiliation{Department of Physics, Massachusetts Institute of Technology, Cambridge, MA 02139}

\author{M.~Greven}
\affiliation{School of Physics and Astronomy, University of Minnesota, Minneapolis, Minnesota 55455, USA}

\author{A.~Lanzara}
\affiliation{Materials Science Division, Lawrence Berkeley National Laboratory, Berkeley, California 94720, USA}
\affiliation{Department of Physics, University of California, Berkeley, California 94720, USA}

\author{J.~Orenstein}
\affiliation{Materials Science Division, Lawrence Berkeley National Laboratory, Berkeley, California 94720, USA}
\affiliation{Department of Physics, University of California, Berkeley, California 94720, USA}

\begin{abstract}
	{The condensation of an electron superfluid from a conventional metallic state at a critical temperature $T_c$ is described well by the BCS theory. In the underdoped copper-oxides, high-temperature superconductivity condenses instead from a nonconventional metallic "pseudogap" phase that exhibits a variety of non-Fermi liquid properties. Recently, it has become clear that a charge density wave (CDW) phase exists within the pseudogap regime, appearing at a temperature $T_{CDW}$ just above $T_c$. The near coincidence of $T_c$ and $T_{CDW}$, as well the coexistence and competition of CDW and superconducting order below $T_c$, suggests that they are intimately related. Here we show that the condensation of the superfluid from this unconventional precursor is reflected in deviations from the predictions of BSC theory regarding the recombination rate of quasiparticles.  We report a detailed investigation of the quasiparticle (QP) recombination lifetime, $\tau_{qp}$, as a function of temperature and magnetic field in underdoped HgBa$_{2}$CuO$_{4+\delta}$ (Hg-1201) and YBa$_{2}$Cu$_{3}$O$_{6+x}$ (YBCO) single crystals by ultrafast time-resolved reflectivity. We find that $\tau_{qp}(T)$ exhibits a local maximum in a small temperature window near $T_c$ that is prominent in underdoped samples with coexisting charge order and vanishes with application of a small magnetic field.  We explain this unusual, non-BCS behavior by positing that $T_c$ marks a transition from phase-fluctuating SC/CDW composite order above to a SC/CDW condensate below. Our results suggest that the superfluid in underdoped cuprates is a condensate of coherently-mixed particle-particle and particle-hole pairs.}
\end{abstract}

\date{\today}

\maketitle

First observed as stripe-like order in the "214" cuprates ~\cite{TranquadaNATURE95}, as checkerboard order in vortex cores~\cite{HoffmanSCIENCE02}, and subsequently at the surface of Bi and Cl based compounds~\cite{HowaldPNAS03,VershininSCIENCE04,WiseNATPHYS08}, the universality of CDW order in the cuprate phase diagram has been established, through NMR~\cite{WuNATURE11} and X-ray scattering~\cite{GhiringhelliSCIENCE12,ChangNATPHYS12,LeTaconNATPHYS14,CominSCIENCE14,daSilvaNetoSCIENCE14,TabisNATCOM14} probes. In YBa$_{2}$Cu$_{3}$O$_{6+x}$ (YBCO) near hole concentration 1/8 application of large magnetic fields stabilizes long-range CDW order at a temperature that approaches $T_c$~\cite{WuNATURE11,leBoeufNATPHYS13}. The near-degeneracy of the characteristic temperatures of CDW and SC phases suggests that these two order parameters are related, as opposed to simply coexisting and competing. Several theoretical works have suggested that the same short-range antiferromagnetic fluctuations drive the formation of CDW and SC states~\cite{EfetovNATPHYS13,SachdevPRL13,DavisPNAS13,HaywardSCIENCE14,WangPRB14}. Moreover, the temperature dependence of the CDW amplitude in YBCO, as determined from X-ray scattering, can be reproduced by a model of fluctuating, multi-component order, of which CDW and SC states are two projections~\cite{HaywardSCIENCE14}.

The basic premise of BCS theory is that the SC condensate is made up of Cooper pairs, which are bound states of two electrons with opposite momenta and spin~\cite{Schrieffer}. Subsequent to BCS, Kohn and Sherrington~\cite{Kohn} showed that a CDW state is likewise a pair condensate, but of electron and holes, whose net momentum determines the wavelength of charge order. Quasiparticles (or broken pairs) are the fundamental excitations of paired condensates such as the SC and CDW states. It is important for what follows to note that, although quasiparticles in SC and CDW states are fermions, they have properties distinct from the quasiparticles that constitute the normal state. SC quasiparticles are phase-coherent linear superpositions of normal state electrons and holes, while CDW quasiparticles are superpositions of electrons (or holes).

Our experiments probe quasiparticle states in underdoped cuprates through time-resolved measurements of their lifetime against recombination, whereby two quasiparticles of opposite spin re-pair and scatter into the condensate. As we discuss below, this scattering rate is sensitive to the phase-coherence of quasiparticle superposition states. To measure the recombination lifetime, we first generate a nonequilibrium quasiparticle population by photoexcitation with an ultrashort optical pulse. We use a low pump fluence so as to probe the linear regime in which the photogenerated quasiparticle population is small compared to its thermal value near $T_c$. The rate of return to equilibrium is measured by resolving the photoinduced change in optical reflectivity, $\Delta R(t)$, as a function of time, $t$, after absorption of the pump pulse. A wealth of experiments, as reviewed for example in~\cite{AverittJPCM02}, have demonstrated that the appearance of a $\Delta R$ signal reflects the opening of a gap (or gaps) at the Fermi level and that its amplitude is proportional to the nonequilibrium quasiparticle population.

In Figure 1 we show $\Delta R(t)/R$ at several temperatures for underdoped samples of Hg-1201 with $T_c = 55$, $71$, and $91$ K. At high temperature we observe a short-lived negative component of $\Delta R(t)$ that is associated with the pseudogap (PG)~\cite{DemsarPRL99A}. With decreasing $T$ a larger amplitude positive component with a much longer lifetime appears and quickly dominates the signal. In cuprates with near-optimal doping this positive, long-lived component appears close to $T_c$~\cite{HanPRL90} and was therefore associated with the onset of superconductivity. However, this association breaks down in underdoped samples in which the positive component is already large at $T_c$ (data at $T_c$ are highlighted in red).

In Figures 2a-f we plot the maximum value of $\Delta R/R(t)$ (which occurs at $\approx$ 1 ps after photoexcitation) as a function of $T$ in underdoped Hg-1201 samples with $T_c$'s ranging from 55 to 91 K. The temperature ($T_{onset}$) at which the positive component of $\Delta R$ appears is indicated by a blue down arrow in each panel. Note that $\Delta R$ continues to increase continuously with further decrease of $T$, without a clear feature at the critical temperature for superconductivity (indicated by red arrows). Both $T_{onset}$ and $T_c$ are plotted as a function of hole concentration, $p$, in Fig. 2g. The onset temperatures of positive $\Delta R$ outline a dome that peaks on the underdoped side of the phase diagram and extends to temperatures 130 K above $T_c$.

Based on a correlations with other probes, we believe that the appearance of the slow, positive component of $\Delta R$ at the temperatures $T_{onset}(p)$ shown in Fig. 2g corresponds to the onset of local CDW order. There is a clear correspondence between the dome of $T_{onset}$ as determined by $\Delta R(T)$ in Hg-1201 and the region of the phase diagram where a CDW is detected in YBCO~\cite{HuckerPRB14,BlancoPRB14}. Although the phase space region of CDW in Hg-1201 is yet to be mapped in as much detail as in YBCO, a CDW has been detected in Hg-1201 samples with $T_c$=71 K ~\cite{TabisNATCOM14} at a temperature (indicated by the green circle in Fig. 2g) coincident with $T_{onset}$. Another correlation linking $\Delta R$ to the CDW in underdoped cuprates is that the positive component of $\Delta R$ in YBCO Ortho-VIII (to be discussed further below) has the same temperature dependence as a zero wavevector vibrational mode that arises from CDW-induced zone-folding~\cite{HintonPRB13}.

We turn now to measurements of the $T$ dependence of the recombination lifetime of quasiparticles for $T<T_{onset}$. The decay curves in Fig. 1 were fit using a function of the form, $\Delta R(t) \propto A e^{-t/\tau_{qp}}-B e^{-t/\tau_r} + C$, where the first term describes QP recombination, the second term accounts for finite risetime and the presence of a negative PG component, and the constant offset $C$ captures a long-lived contribution that we attribute to local heating by the pump pulse (see Supplement for details on the fitting procedure). Figure 3a displays the evolution of $\tau_{qp}(T)$ with hole concentration in the Hg-1201 system. At each hole concentration we observe structure in the $T$-dependence of the quasiparticle recombination time at $T_c$. In underdoped samples there is a peak in $\tau_{qp}(T)$ at $T_c$ that is most prominent in the $T_c=$71 K sample and decreases in amplitude at lower and slightly higher hole concentration in a manner that appears to be correlated with the strength of the CDW. 

Figs. 3b and 3c compare $\Delta R(t,T)$ in Hg-1201 ($T_c=71$ K) and YBCO Ortho-VIII ($T_c = 67$ K), illustrating that generality of the phenomena described above. Fig. 3b and 3c show the temperature dependence of the amplitude $\Delta R_{qp}\equiv A$ and $\tau_{qp}$ respectively for the two underdoped samples. In both we observe the onset of positive $\Delta R(T)$ well above $T_c$ and a smooth variation through the SC transition. As shown in Fig. 3c, peak in $\tau_{qp}(T)$ centered on $T_c$ is strikingly similar in the two representative samples.

The $T$-dependence of the quasiparticle lifetime in the sample of Hg-1201 at near-optimal doping (topmost data set of Fig. 3a) is qualitatively different from what is seen in underdoped samples. In the near-optimal sample $\tau_{qp}$ grows monotonically as $T \rightarrow T_c$, suggesting a tendency to diverge as $\Delta R_{qp}(T)$ goes to zero, while in underdoped samples the peak in $\tau_{qp}$ at $T_c$ is a small feature on a smooth background. The behavior of $\tau_{qp}$ in near-optimal Hg-1201, which is observed in other optimally doped cuprates as well~\cite{HanPRL90,KabanovPRB99}, can be understood within the mean-field theory of superconductivity. The theoretical description of $\tau_{qp}(T)$ in the context of BCS began in the 1960's and its subsequent history is reviewed in Ref.~\cite{Schuller}. In the mean-field picture, the relaxation of a nonequilibrium quasiparticle population is described by a pair of coupled equations: a Landau-Khalatnikov equation for the energy gap and a Boltzmann-like equation that governs the quasiparticle distribution~\cite{Schuller}. According this analysis, $\tau_{qp}(T)\propto\tau T/\Delta(T)$ (where $\tau$ is the electron inelastic scattering time) and diverges as the superconducting gap $\Delta(T)$ vanishes as $T$ approaches $T_c$ from below. In more recent work on cuprate superconductors, effects associated with the phonon bottleneck~\cite{RT,KabanovPRB99,GedikPRB04} are included, which leads to replacing $\tau$ by the lifetime of phonons with energy greater than $2\Delta$.

The behavior of $\tau_{qp}(T)$ in underdoped samples is at odds with the mean-field picture described above. Two observations -- continuous variation of $\Delta R_{qp}$ for $T<T_{onset}$ and the smoothly varying background in $\tau_{qp}(T)$ that underlies the small peak -- suggest that a quasiparticle gap has opened well above $T_c$. Given a pre-existing gap, it is very difficult to explain the modulation of the quasiparticle lifetime near $T_c$. In view of the difficulties with the mean-field picture, we are led to consider the onset of phase coherence at $T_c$, rather than gap-opening, as the origin of the structure in $\tau_{qp}(T)$.

Coherence effects have been observed previously elastic QP scattering, as detected by QP interference in scanning tunneling microscopy experiments~\cite{HoffmanSCIENCE02,HanaguriSCIENCE09,LeeSCIENCE09}. However, the implications of coherence for a process in which two quasiparticles scatter into the condensate have not previously been considered in the context of the cuprates. In a phase-incoherent state the recombination rate is proportional to the square modulus of the interaction matrix element between electrons and holes. In a phase-coherent paired state, recombination is more complex because, as mentioned previously, QPs are linear superpositions of electrons and holes. As a result, the matrix element for recombination reflects multiple channels that can interfere constructively or destructively, depending on the nature of the pairing state and QP interactions~\cite{Schrieffer}.

The rate of QP recombination in a paired condensate is the product of the normal state rate and a "coherence factor" ($F$) that is a function of the Bogoliubov coefficients $u$ and $v$. For the case of time-reversal invariant QP interactions the coherence factors for SC and CDW condensates are given by
\begin{equation}
F_{SC} = (v u^{\prime}- u v^{\prime})^{2} = \frac{1}{2}\left(1 + \frac{\Delta \Delta^{\prime}}{E E^{\prime}}\right)
\end{equation}
\begin{equation}
F_{CDW} = \frac{1}{2}\left(1 - \frac{\Phi \Phi^{\prime}}{E E^{\prime}}\right)
\end{equation}
where $\Delta$ and $\Delta^{\prime}$ are the amplitudes of the SC gap at the two QP momenta, $\Phi$ and $\Phi^{\prime}$ are the analogous CDW gaps, and $E$ and $E^{\prime}$ are the QP energies~\cite{Gruner}. In the limit that $E$ and $E^{\prime}$ approach the gap energy, these factors reduce to $F_{SC} = 1$ and $F_{CDW} = 0$, while in the normal state $F=1/2$. We note that the result $F_{CDW}=0$ is related to the $\pi$ phase shift between occupied and unoccupied states, which has recently been observed in CDW states in Bi2212~\cite{HamidianNATPHYS15}.

The simplest scenario, in which a phase-fluctuating superconductor becomes fully phase coherent at $T_c$, is inconsistent with a peak in $\tau_{qp}(T)$ (or local minimum in recombination rate). Instead, SC coherence yields a doubling of the recombination rate, corresponding to the factor of two jump in $F$ from 1/2 to 1 upon crossing from a normal to SC state. However, we have found that a model that takes into account the dual presence of fluctuating CDW and SC order leads to a singular feature in $\tau_{qp}(T)$ that agrees well with experiment.

To formulate this model quantitatively, we derived the recombination coherence factor for a state with coexisting SC and CDW order. The composite SC/CDW condensate is made of particle-hole quadruplets, pairing electrons at $\pm k$ and holes at $\pm (k+Q)$ separated by the CDW wavevector $Q$. The structure of the quasiparticle eigenstates of the composite SC/CDW condensate is shown schematically in Fig. 4a. Based on these eigenstates, we determined the dependence of the mixed state coherence factors on the quasiparticle energy (see Methods).  As time and angle-resolved photoemission measurements observe rapid thermalization of quasiparticles to gap edge after photo-excitation~\cite{SmallwoodSCIENCE12,SmallwoodPRB15}, we focus on the coherence factor in the limit that $E, E^{\prime}\rightarrow 0$, which simplifies the expression for $F$ considerably. In this limit, the QP recombination coherence factor becomes,
\begin{equation}
F = \frac{1}{2}\left(1 + \cos\phi_{\Delta}\frac{\Delta^{2}}{\Delta^{2}+\Phi^{2}}-\cos\phi_{\Phi}\frac{\Phi^{2}}{\Delta^{2}+\Phi^{2}}\right)
\end{equation}
where $\phi_\Delta$ and $\phi_\Phi$ are the relative phases of the QP pair undergoing recombination. In the case of $d$-wave pairing, these phases will be $0$ or $\pi$, depending on which $k$ points the QPs occupy. STM QP interference demonstrates that the strongest scattering channels are between states with the same sign of gap amplitude~\cite{LeeSCIENCE09}, so we restrict our attention to these recombination processes.

In the presence of phase fluctuations, the $\cos\phi$ factors are replaced by their ensemble average $\left<\cos\phi\right> = e^{-\tau_0/\tau_c}$, where $\tau_{c}$ is the phase-correlation time and $\tau_0$ is the QP lifetime in the fully incoherent regime.  The coherence factor obtained by this substitution would apply to systems with coexisting, but independent, CDW and SC order. However, in the light of evidence that these orders are strongly coupled in underdoped cuprates, we consider a description in terms of a multi-component order parameter~\cite{HaywardSCIENCE14} whose amplitude $\sqrt{\Phi^{2}+\Delta^{2}}$ is constant and whose fluctuations are described by a single phase $\phi$. With these assumptions the coherence factor can be written,

\begin{equation}
\left<F\right> = \frac{1}{2}\left(1-e^{-\tau_0/\tau_c}\cos2\theta\right),
\end{equation}
where $\tan^2\theta\equiv \Phi^2/\Delta^2$.

Fig. 4d shows a fit to $\tau_{qp}(T)$ for the Hg-1201 $T_c=71$ K sample using Eq. 4. Figs. 4b and 4c show the temperature dependence of the three parameters from which the fits were generated. The mixing angle $\theta(T)$ is plotted in Fig. 4b.  The reciprocal of the coherence time, $\tau_c^{-1}$, and reciprocal of the incoherent recombination time, $\tau_0^{-1}(T)$, are plotted as solid and dashed lines, respectively in Fig. 4c. The $T$-dependence of $\tau_0$ is determined by a polynomial fit to $\tau_{qp}(T)$ that ignores the peak at $T_c$. The quasiparticle decoherence rate $\tau_c^{-1}$ is constrained to be constant below $T_c$ and to vary $\propto (T-T_c)$ above the transition, as suggested by angle-resolved photoemission spectra~\cite{NormanPRB98}. The slope of $\tau_c^{-1}$ vs. $T-T_c$ extracted from our fit ($\approx 0.15$ THz/K) is consistent with estimates of coherence times obtained from optical conductivity~\cite{Corson}. The peak in $\tau_{qp}(T)$ shown in Fig. 4d arises from the interplay between the mixing angle $\theta(T)$ and the onset of quasiparticle coherence. Starting at temperatures well above $T_c$, the order parameter is CDW-like with a short coherence time. As the temperature is lowered towards $T_c$ and the CDW becomes more coherent, $\left<F\right>$ dips below its normal state value of $1/2$ and the QP lifetime is enhanced. However, as the order parameter crosses over from CDW to SC-like near $T_c$, $\left<F\right>$ increases, giving rise to a peak $\tau_{qp}$.

To further examine the relationship between phase coherence and quasiparticle recombination, we investigated the effect of magnetic field, $B$, on $\tau_{qp}(T)$. An overview of $\tau_{qp}(B,T)$ in fields applied perpendicular to the Cu-O planes is shown in Figure 5. Fig. 5a compares the quasiparticle lifetime in zero field and 6 Tesla for a sample of Hg-1201 with $T_c=$ 71 K. The peak in $\tau_{qp}(T)$ is entirely washed out by the field and replaced by a smooth background that is described by the model parameter $\tau_0(T)$ discussed previously. In Fig. 5c, the change in recombination rate caused by the field, $\Delta\gamma_{qp}(B)\equiv \tau_{qp}^{-1}(B)-\tau_{qp}^{-1}(0)$, is plotted vs. $B$ at three temperatures: well below, above, and at $T_{c}$. It is clear that strong dependence of $\tau_{qp}$ on $B$ is observed only near $T_c$.  The field dependence of the quasiparticle lifetime is qualitatively different in the near optimal Hg-1201 sample (shown in Fig. 5b), where we find that the maximum $\tau_{qp}$ shifts to lower $T$ but is not reduced, consistent with what is expected for a mean-field gap opening transition.

The $B$ dependence of $\tau_{qp}$ can be understood to be a consequence of dephasing in the vortex liquid induced by the field. Assuming a total dephasing rate, $\tau_c^{-1}+\Gamma(B)$, where $\Gamma(B)$ is the dephasing rate associated with vortex diffusion, yields

\begin{equation}
\Delta\gamma_{qp}(B)=\left[\tau_{qp}^{-1}(T_c)-\tau_0^{-1}(T_c)\right/2] \{1-\exp\left[-\Gamma(B)\tau_0\right]\}.
\end{equation}
The dashed line in Fig. 5c is a fit to this functional form, with $\Gamma(B)=$(0.08 THz/T)$B$. This result can be compared with the theory of phase fluctuations in the vortex liquid~\cite{HalperinNelson}, where it is shown that $\Gamma(B)\simeq (\xi/l_B)^2k_BT_c/\hbar$, where $\xi$ and $l_B$ are the coherence and magnetic length, respectively. Equating this estimate with the measured $\Gamma(B)$ yields a reasonable value for the coherence length of 2.4 nm, strongly suggesting that $B$-induced dephasing accounts for wipeout of the $\tau_{qp}$ peak.

To summarize, we have described measurements and analysis of the photoinduced transient reflectivity, $\Delta R(t,T)$ in representative YBCO and Hg-1201 samples of underdoped cuprates. The onset of $\Delta R$ with decreasing temperature is correlated with the appearance of the incommensurate CDW detected previously by NMR, STM, and X-ray scattering measurements.  This correlation leads us to conclude that, although the density of states depression known as the pseudogap is formed at a higher temperature, the CDW either enhances it, or leads to a new gap of different origin. A correlation between CDW and gap formation is suggested as well by recent ARPES measurements~\cite{HashimotoNATPHYS10,HeSCIENCE11}.  We focused attention on two aspects of $\Delta R(t)$ as $T$ was lowered through $T_c$.  In underdoped samples, $\Delta R$ increases continuously through $T_c$, suggesting a smooth variation of the gap at the Fermi surface hot spots (or Fermi-arc tips). Second, while the gap varies continuously, the quasiparticle recombination time exhibits a narrow local maximum at $T_c$.  We proposed that this peak in $\tau_{qp}$ indicates a crossover from fluctuating CDW to SC/CDW order, occurring as the condensate coherence time slides through the time window set by the background recombination time $\sim$ 2-3 ps.  The link between condensate coherence and the peak in $\tau_{qp}(T)$ was further supported by the observation that magnetic field causes the peak to disappear into the background. Our results show that quasiparticle recombination provides a new method for probing the onset of coherence in systems characterized by a fluctuating multi-component order parameter.

\textbf{Acknowledgements:} Synthesis and characterization of Hg-1201 samples performed at the University of Minnesota was supported by the Department of Energy, Office of Basic Energy Sciences, under Award No. DE-SC0006858. N.B. acknowledges the support of FWF project P2798. Optical measurements and modeling performed at Lawrence Berkeley National Lab was supported by the Director, Office of Science, Office of Basic Energy Sciences, Materials Sciences and Engineering Division, of the U.S. Department of Energy under Contract No. DE-AC02-05CH11231.

\section{Materials and Methods}

\subsection{Hg-1201 synthesis and preparation}
We examined a series of high quality single crystals of Hg-1201 ranging from deeply underdoped to optimal doping with $T_{c}$'s of 55 K, 65 K, 71 K, 79 K, 81 K, 91 K, and 94 K. These samples were grown by the self-flux method~\cite{Zhao} and annealed to achieve different oxygen concentrations~\cite{Barisic}, with $T_{c}$ determined within $\sim 1$ K by magnetic susceptibility measurements. Hole concentration was determined using methods described in Ref.~\cite{Yamamoto}. Samples were polished under nitrogen flow using 0.3 $\mu$m grit films to prevent surface oxidation.

\subsection{Ultrafast Measurements}
The measurements reported herein were performed using 100 fs pulses from a mode-locked Ti:Sapphire laser at 800 nm center wavelength and $\simeq1 \mu$J$/$cm$^2$ fluence. Pump pulses induced a small ($\sim10^{-4}$) fractional change in reflectivity that was monitored by time-delayed probe pulses. At the low laser fluence used in this study, samples are weakly perturbed by the pump pulse, in the sense that the density of photogenerated quasiparticles is much less than the thermal equilibrium value. In this regime the decay rate is a measure of the quasiparticle recombination rate in thermal equilibrium. Measurements in zero magnetic field at Berkeley were performed under vacuum in an Oxford continuous-flow liquid He cryostat. Magneto-optic measurements at Berkeley were performed in a 6T Oxford Spectramag cryostat, and those at MIT in a 7T Janis cryostat.

\subsection{Calculation of mixed-state coherence factors}
The mean-field Hamiltonian is given by $\mathcal{H}_{SC+CDW} = \Sigma_{\mathbf{k}}A^{\dag}_{\mathbf{k}} M A_{\mathbf{k}}$~\cite{LevinPRB74}, with
\begin{align}
A^{\dag}_{\mathbf{k}} = &\left(\begin{smallmatrix}a^{\dag}_{\mathbf{k} \uparrow} & a^{\dag}_{(\mathbf{k}\pm Q) \uparrow} & a_{-\mathbf{k} \downarrow} & a_{-(\mathbf{k}\pm Q) \downarrow}\end{smallmatrix}\right)\\
M =& \left(\begin{smallmatrix}
\xi_{\mathbf{k}} & \Phi & \Delta & 0 \\
\Phi^\star & \xi_{\mathbf{k}\pm Q} & 0 & \Delta \\
\Delta^\star & 0 & -\xi_{-\mathbf{k}} & -\Phi\\
0 & \Delta^\star & -\Phi^\star & -\xi_{-(\mathbf{k}\pm Q)}
\end{smallmatrix}\right)
\end{align}
Diagonalizing the CDW and SC subsystems via successive Bogoliubov transformations yields energy eigenvalues $E^{2}_{\mathbf{k}} = \pm\left(\xi^{2}_\mathbf{k}+\Delta^{2}+\Phi^{2}\right)$. The eigenvectors give the composite QP operators $\Gamma^{\dag}_{\mathbf{k}} = A^{\dag}_{\mathbf{k}} B$, where
\begin{align}
\Gamma^{\dag}_{\mathbf{k}} = &\left(\begin{smallmatrix}\gamma^{\dag}_{\mathbf{k} \alpha 0} & \gamma^{}_{\mathbf{k} \beta 1} & \gamma^{}_{\mathbf{k} \beta 0} & \gamma^{\dag}_{\mathbf{k} \alpha 1}\end{smallmatrix}\right)\\
B =& \left(\begin{smallmatrix}
s u & t u & s v & t v \\
-t^\star u & s u & t^\star v & -s v \\
-s v^\star & -t v^\star & s u & t u \\
-t^\star v^\star & s v^\star & -t^\star u & s u
\end{smallmatrix}\right)\\
\lvert s_{\mathbf{k}}\rvert^{2} =& 1-\lvert t_{\mathbf{k}}\rvert^{2} = \frac{1}{2}\left(1+\frac{\xi_{\mathbf{k}}}{\sqrt{\xi^2_{\mathbf{k}}+\Phi^{2}}}\right)\\
\lvert u_{\mathbf{k}}\rvert^{2} =& 1-\lvert v_{\mathbf{k}}\rvert^{2} = \frac{1}{2}\left(1+\frac{\sqrt{\xi^2_{\mathbf{k}}+\Phi^{2}}}{\sqrt{\xi^2_{\mathbf{k}}+\Phi^{2}+\Delta^{2}}}\right)
\end{align}
In the above, $u$ and $s$ are assumed to be real, and the phases of $v$ and $t$ are determined by the phases of $\Delta$ and $\Phi$, respectively. We calculate the coherence factors following Ref.~\cite{Tinkham}, starting with a quasiparticle interaction of the form $\mathcal{H}_{int} = \Sigma_{\mathbf{k}} I_{\mathbf{k}^{\prime} \sigma^{\prime} \mathbf{k} \sigma} a^{\dag}_{\mathbf{k}^{\prime} \sigma^{\prime}} a^{}_{\mathbf{k} \sigma}$. We assume that the interaction is spin-independent. The four four electronic transitions that involve the same QP operators,
\begin{align}
& a^{\dag}_{\mathbf{k}^{\prime} \uparrow} a^{}_{\mathbf{k} \uparrow} + a^{\dag}_{(\mathbf{k}^{\prime}\pm \mathbf{Q}) \uparrow} a^{}_{(\mathbf{k}\pm\mathbf{Q}) \uparrow}\\
+ & a^{\dag}_{-\mathbf{k} \downarrow} a^{}_{-\mathbf{k}^{\prime} \downarrow} + a^{\dag}_{-(\mathbf{k}\pm \mathbf{Q}) \uparrow} a^{}_{-(\mathbf{k}^{\prime}\pm\mathbf{Q}) \uparrow}
\end{align}
must be summed coherently before calculating the squared modulus of the matrix element that determines the quasiparticle scattering rate. Expressing the four bilinear operators in terms of the quasiparticle ($\gamma$) operators yields,
\begin{align}
& f_{1}\gamma_{\mathbf{k} \alpha 0} \gamma_{\mathbf{k}^{\prime} \beta 0} +
f^{\star}_{1}\gamma_{\mathbf{k} \alpha 1} \gamma_{\mathbf{k}^{\prime} \beta 1}\\
+ & f_{2}\gamma_{\mathbf{k} \alpha 0} \gamma_{\mathbf{k}^{\prime} \beta1} +
f^{\star}_{2}\gamma_{\mathbf{k} \alpha 1} \gamma_{\mathbf{k}^{\prime} \beta 0}
\end{align}
where,
\begin{align}
f_{1} = & u s v^\prime s^{\prime}+u t v^{\prime} t^{\prime \star}+v s u^{\prime} s^{\prime}+v t u^{\prime} t^{\prime \star}\\
f_{2} = & u t^{\star} u^{\prime} s^{\prime}-u s u^{\prime} t^{\prime \star}+v t^{\star} v^{\prime} s^{\prime}-v^{\star} s v^{\prime} t^{\prime \star}
\end{align}
and $u^{\prime} \equiv u_{\mathbf{k}^{\prime}}$. Summing over recombination channels and then squaring yields a coherence factor for quasiparticle recombination given by $F = \lvert{f_{1}}\rvert^{2}+\lvert{f_{2}}\rvert^{2}$.

\begin{figure}[H]
	\begin{center}
		\includegraphics[width=17cm]{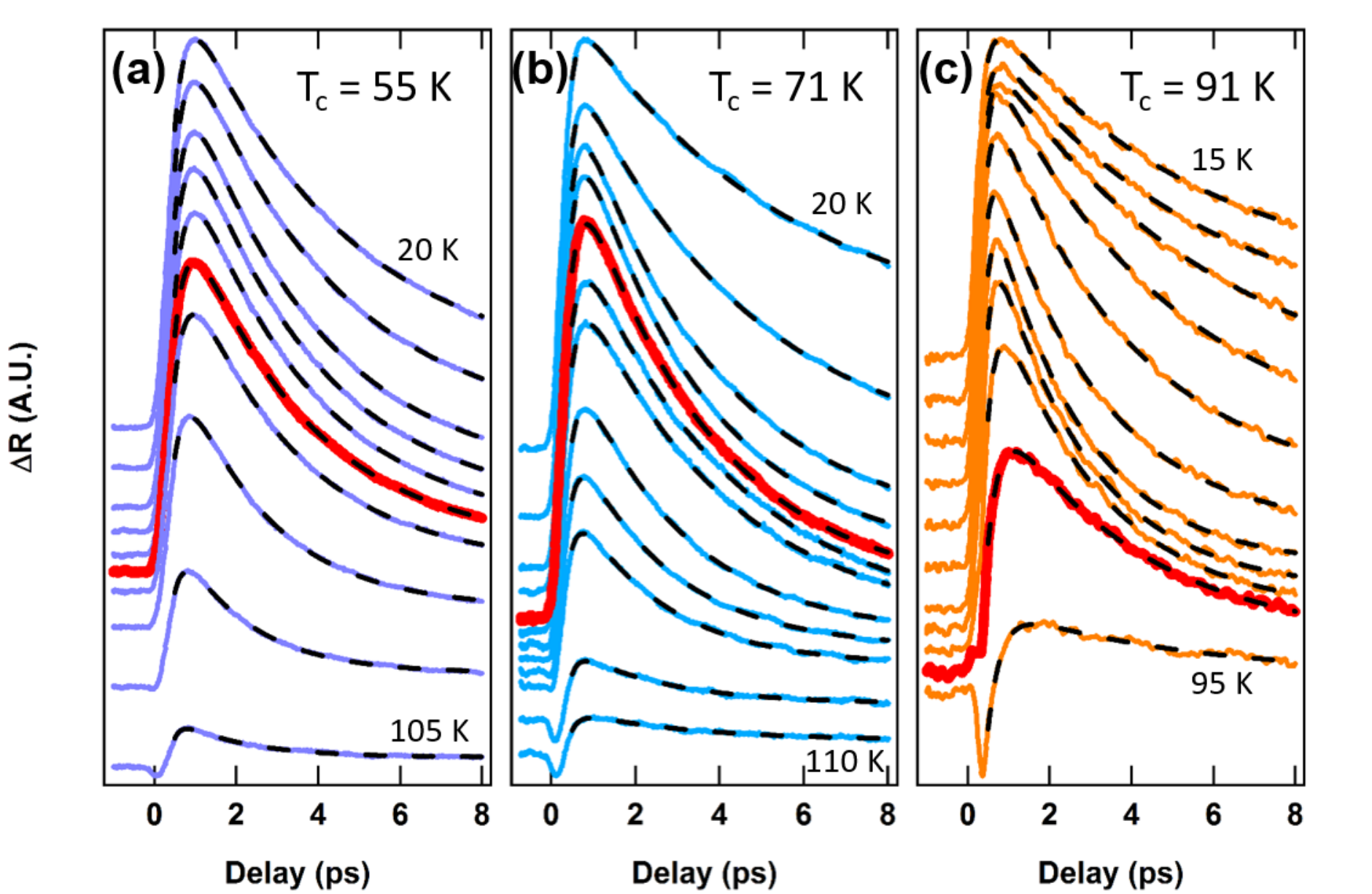}
	\end{center}
	\caption{\label{fig:1} Time-dependence of $\Delta R(t,T)/R$ for a range of temperatures that spans $T_c$ is shown for three underdoped samples with $T_c = 55$, $71$, and $91$ K, in (a), (b), and (c), respectively. Curves are offset for clarity by an ammount proportional to the temperature. The quasiparticle recombination time is extracted for each temperature from the exponential fits shown as dashed black lines.}
\end{figure}

\begin{figure}[H]
	\begin{center}
		\includegraphics[width=17cm]{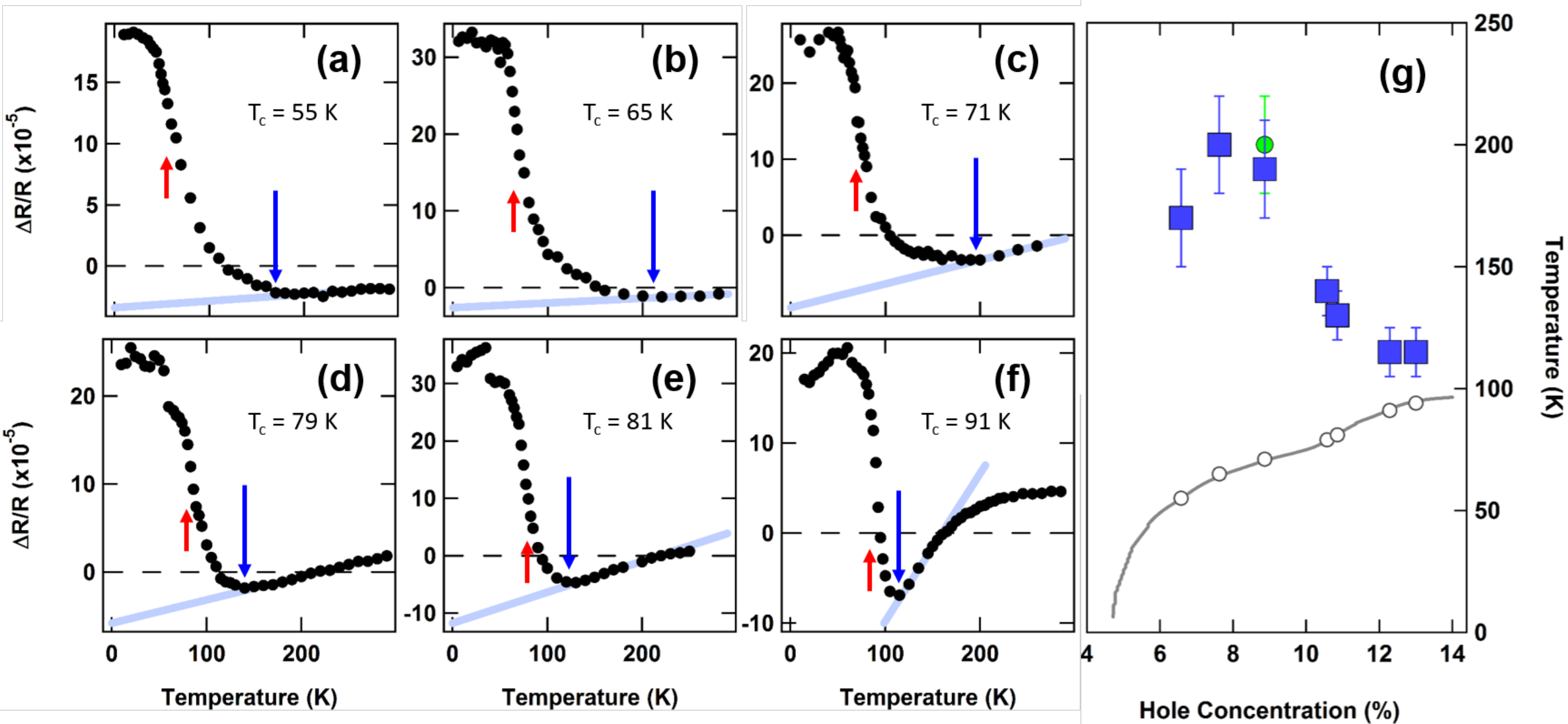}
	\end{center}
	\caption{\label{fig:2}(a)-(f) The temperature dependence of $\Delta R(t = 1$ps$)/R$ is plotted for a series of doping levels. Values of $T_{onset}$ as determined from the inflection points of the curves are indicated by the vertical blue arrows, and $T_{c}$ is indicated by red arrows. (g) $T_{onset}$ as function of hole concentration, $p$, as obtained from the inflection points in Figs. (a)-(f) are plotted as blue squares in a $T-p$ phase diagram for the Hg-1201 system. Also shown is the onset temperature of charge order observed by X-ray scattering~\cite{TabisNATCOM14} (green circle) and the critical temperatures for superconductivity (open circles).}
\end{figure}

\begin{figure}[H]
	\begin{center}
		\includegraphics[width=17cm]{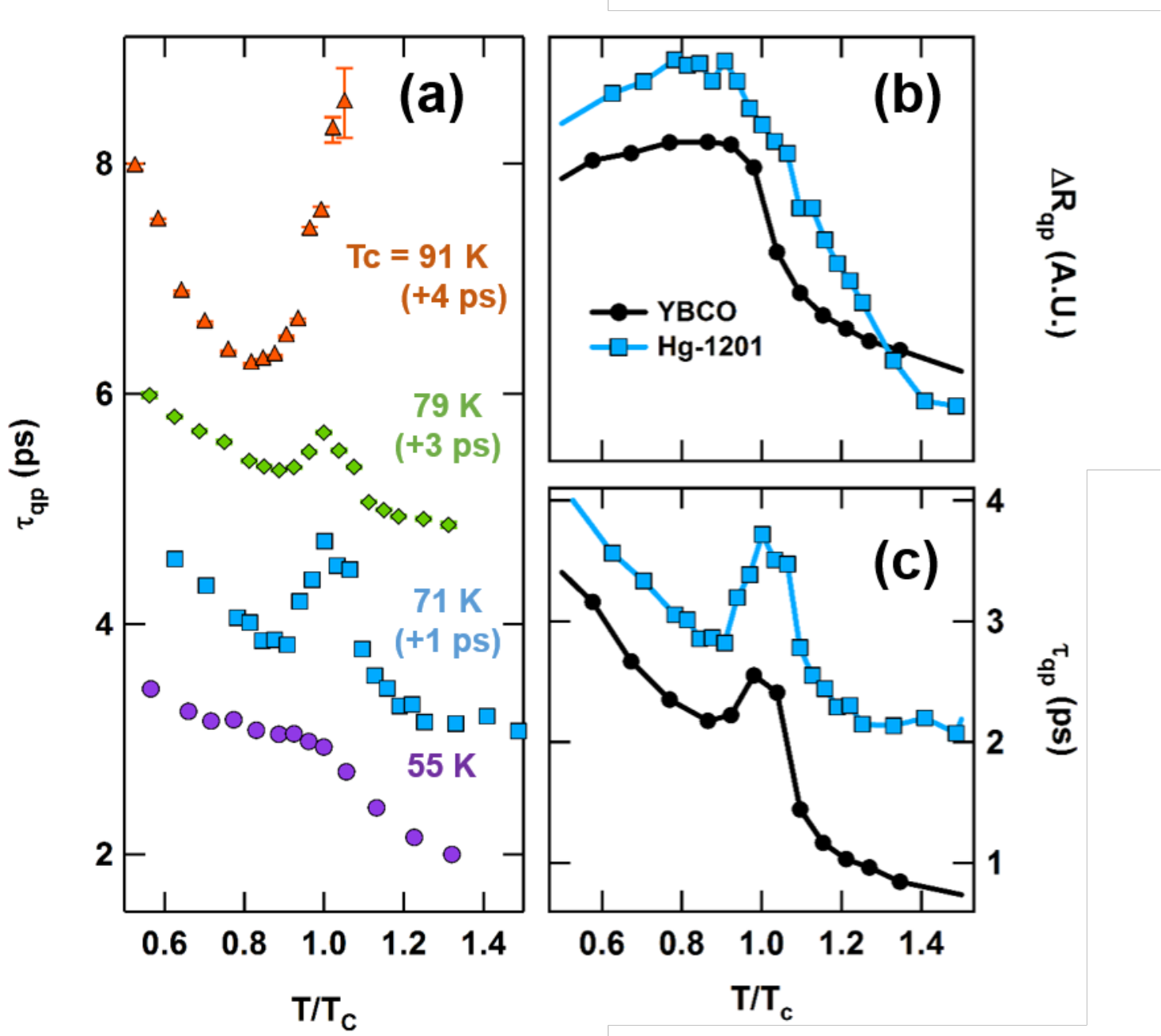}
	\end{center}
	\caption{\label{fig:3}(a) Temperature dependence of $\tau_{qp}$ for a series of underdoped Hg-1201 samples. Curves are offset for clarity, as indicated in the figure. (b) Temperature dependence of the positive component of the transient reflectivity, $\Delta R_{QP}$ and (c) the quasiparticle lifetime, $\tau_{qp}$, for YBCO Ortho VIII with $T_c = 67$ K (black circles) and Hg-1201 with $T_c = 71$ K (blue squares). The temperature axis is normalized to $T_c$.}
\end{figure}

\begin{figure}[H]
	\begin{center}
		\includegraphics[width=17cm]{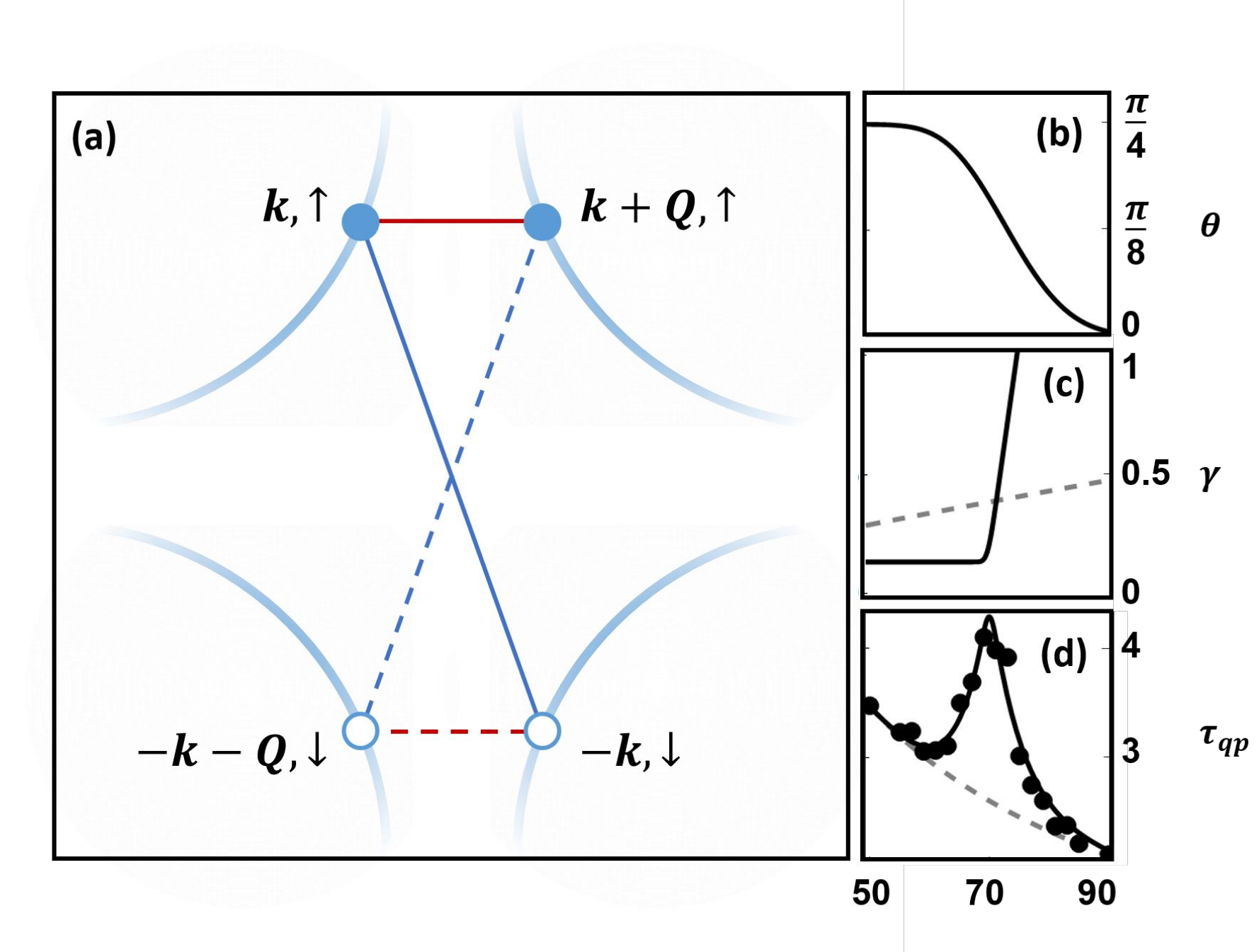}
	\end{center}
	\caption{\label{fig:4} (a) Momentum space diagram of the electron eigenstates that determine the rate of quasiparticle recombination.  The region defined by the square is the first Brillouin zone of a cuprate superconductor - inside are the four "Fermi arcs" shown as blue curves.  The pairing of electrons and holes at $k$ and $k+Q$ drives CDW formation, whereas electron pairing at $k$, $-k$ states (blue lines) is responsible for superconductivity. Quasiparticles are excited states of paired condensates formed from a superposition of two normal phase electron eigenstates.  CDW quasiparticles, represented by the red lines, are superpositions of electrons at $k$ and $k+Q$, whereas SC quasiparticles, represented by blue lines, are superpositions of electrons and holes.  In the mixed SC-CDW condensate that forms in underdoped cuprate superconductors, quasiparticles become superpositions of four, rather than just two electron eigenstates.  The rate at which such quasiparticles scatter back into the condensate depends sensitively on the mixing angle of SC and CDW components, as well as on the condensate phase-correlation time. The temperature dependent mixing angle, $\theta(T)$ (shown in (b)), phase fluctuation rate, $\tau_c^{-1}(T)$ (solid line in (c)), and reciprocal of the normal state lifetime, $\tau_0^{-1}(T)$ (gray dashed line in (c)) produce the fit to the quasiparticle lifetime in the Hg-1201 ($T_c=71$ K) sample displayed in (d).}
\end{figure}

\begin{figure}[H]
	\begin{center}
		\includegraphics[width=17cm]{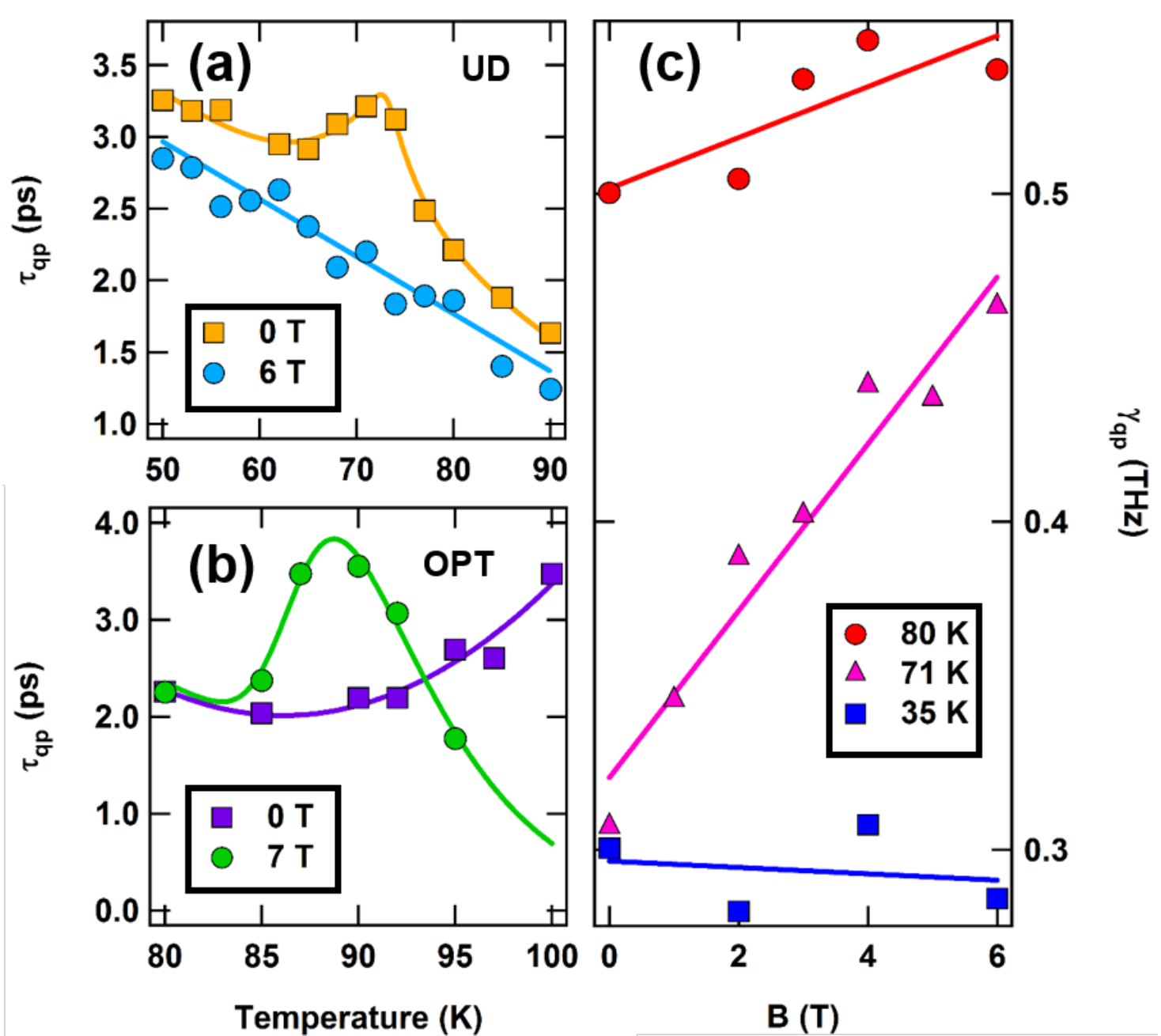}
	\end{center}
	\caption{\label{fig:5}(a) Temperature dependence of the recombination time $\tau_{qp}$ in zero field and 6 Tesla applied normal to the surface of Hg-1201 ($T_c=71$ K). (b) Temperature dependence of $\tau_{qp}$ at 0 and 7 Tesla in Hg-1201 ($T_c=94$) K. (c) The quasiparticle recombination rate, $\gamma_{qp}$, at 35 K, at $T_c$ = 71 K, and at 80 K, plotted vs $H$ for the underdoped Hg-1201 sample with $T_c=71$ K.}
\end{figure}

\pagebreak

\section{Supporting information}

\subsection{Fitting procedure}

Here we describe details of the fits to $\Delta R(t)/R$ to demonstrate that the values of $\tau_{qp}(T)$ discussed in the main text are independent of the fitting procedure. Fig. S1 (a) shows time-resolved reflectivity data for five temperatures below, at, and above $T_c$ for the Hg-1201 sample with $T_c = 71$ K. The three-component fits, displayed in red, are of the form described in the text, $\Delta R = A e^{-t/\tau_{qp}}- B e^{-t/\tau_r}+C$. Fig S1 (b) and (c) display the temperature dependence of the parameters extracted from these fits. For each temperature, the fit extends from the delay at which $\Delta R(t)$ first attains 50\% of its peak value to $t=8$ ps. The parameters $B$ and $\tau_r$ account for both the finite rise time of $\Delta R$ and the presence of the negative PG component. These two contributions cannot be easily disentangled, but they can be captured in the fit by a single negative term in $\Delta R$.  As can be clearly seen, the structure in $\tau_{qp}$ near $T_c$ is not present in the other fit parameters.  This confirms that the peak in $\tau_{qp}(T)$ is representative of the rate of quasiparticle recombination and is not related to structure in the other parameters. The time-independent component $C$ reflects the long-lived photo-induced heating of the system and the non-exponential decay that occurs at low temperatures. This component is much smaller than the exponential components and does not affect the other fit parameters.

In order to further demonstrate that the structure reported in $\tau_{qp}$ is not related to the parameters $B$ and $\tau_r$, we can limit the fit to an exponential decay plus constant, for example by considering $\Delta R$ only after it has decreased to 75$\%$ of its peak value. Fits to a two-component form $\Delta R = A e^{-t/\tau_{qp}}+C$ are displayed in Fig. S2 (a) and the resulting fit parameters are compared to those from the three component fits in Fig. S2 (b)-(c). It is clear that the quantities of interest, $A$ and $\tau_{qp}$, are not affected by the details of the fit procedure.

\begin{figure}[H]
	\begin{center}
		\includegraphics[width=17cm]{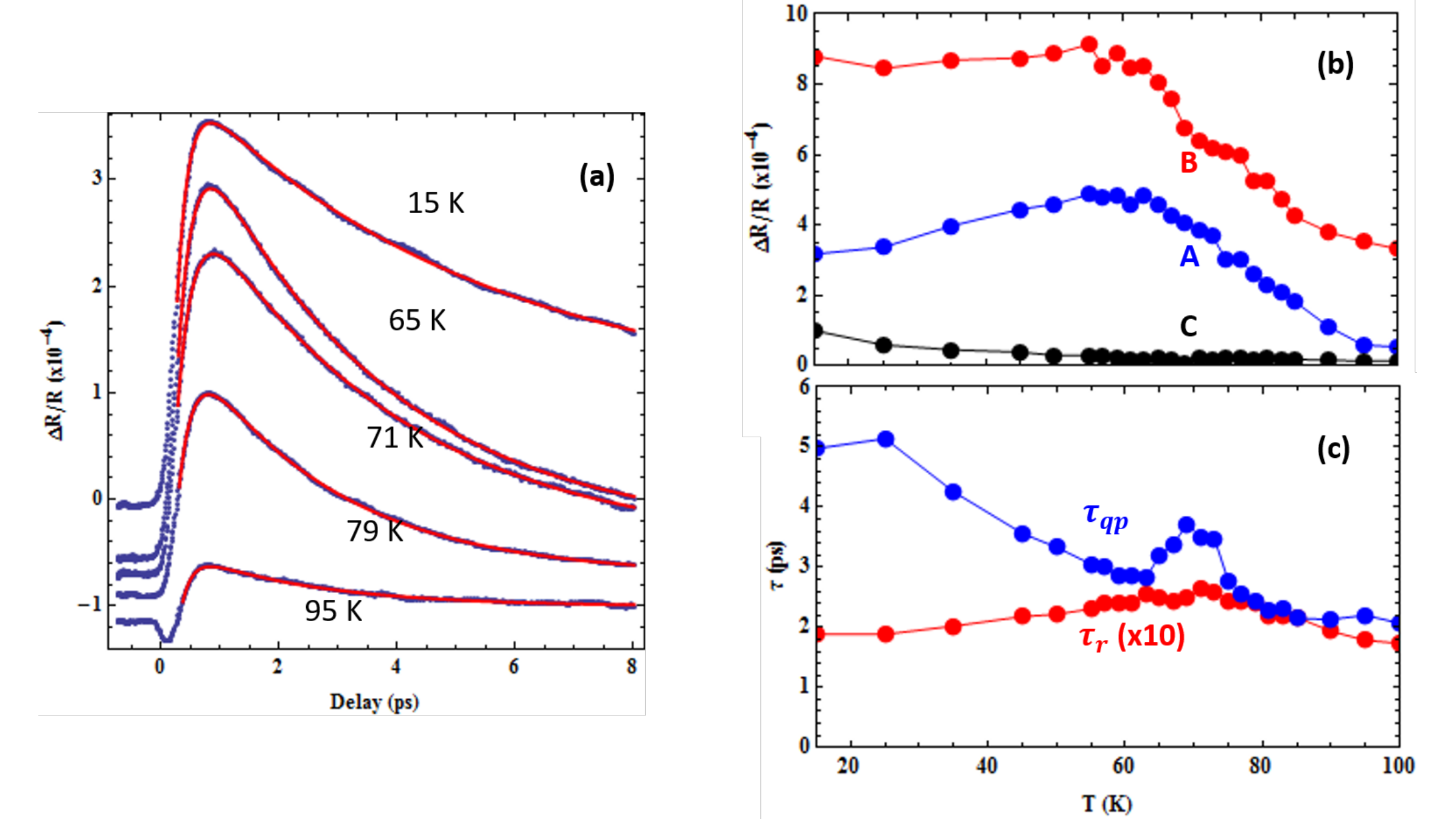}
	\end{center}
	\caption{\label{fig:S1} (a) Select $\Delta R/R$ curves from the $T_c = 71$ K sample of Hg1201. The three-component fits are shown in red. (b) Temperature dependence of the three amplitudes $A$, $B$, and $C$. (c) Temperature dependence of the rise and decay time constants $\tau_r$ and $\tau_{QP}$. $\tau_r$ is multiplied by 10 for easy comparison.}
\end{figure}
	
\begin{figure}[H]
	\begin{center}
		\includegraphics[width=17cm]{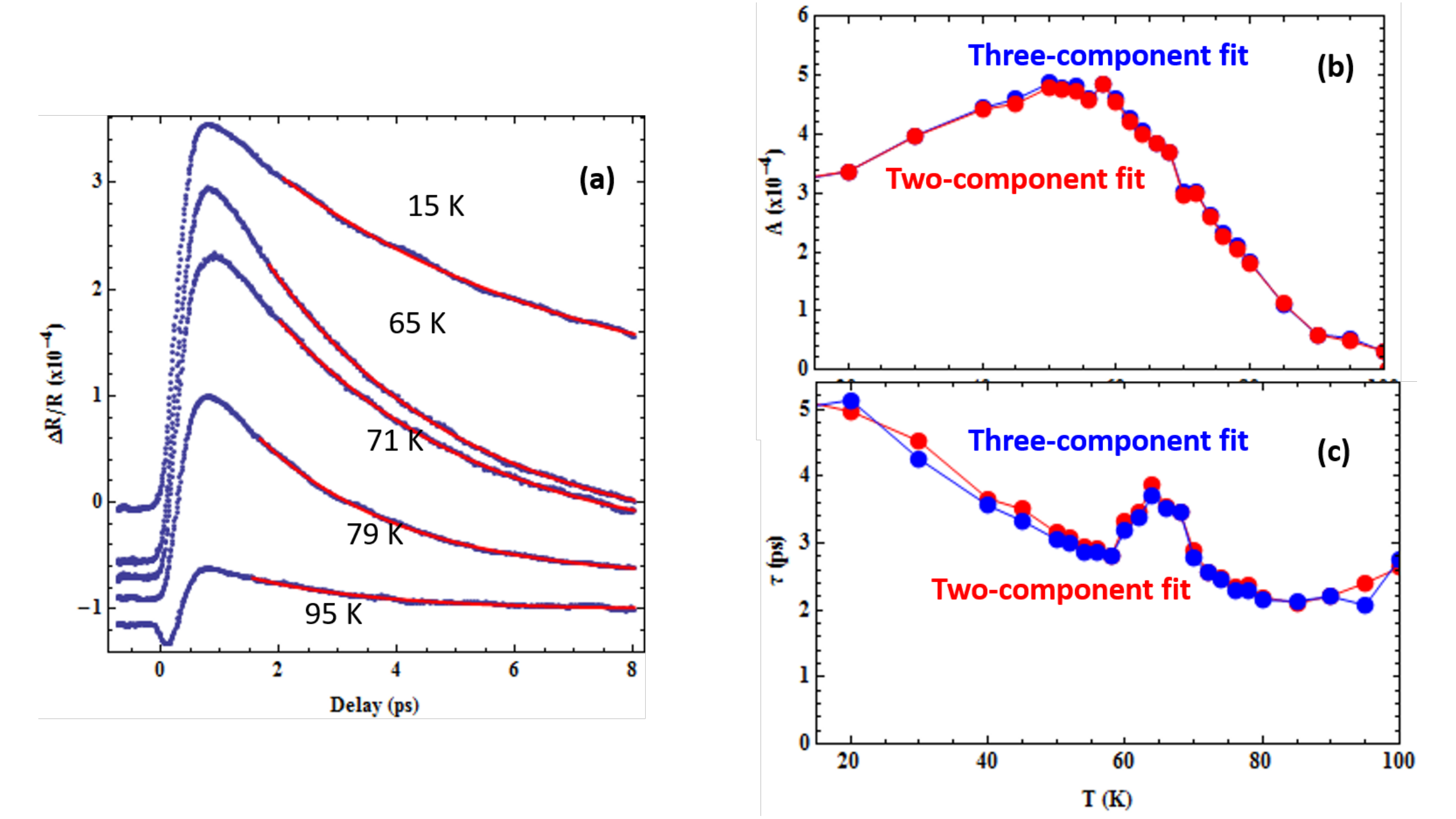}
	\end{center}
	\caption{\label{fig:S2}(a) Select $\Delta R/R$ curves from the $T_c = 71$ K sample of Hg1201. The two-component fits are shown in red. (b) Comparison between $B(T)$ extracted from the three-component fit (blue) and the two-component fit (red).  (c) Comparison between $\tau_{QP}(T)$ extracted from the three-component fit (blue) and the two-component fit (red).}
\end{figure}

\end{document}